\documentclass[reprint,aps,prl,floatfix,twocolumn,superscriptaddress,showpacs,amsmath,amssymb,showpacs,]{revtex4-1} 

\usepackage{graphicx}
\usepackage{epstopdf}
\usepackage{epsfig}
\usepackage{epsf}
\usepackage{url}
\usepackage[USenglish]{babel}
\usepackage{hyperref}
\def\bcen{\begin{center}}
\def\ecen{\end{center}}
\allowdisplaybreaks
\renewcommand\[{\begin{equation}}
\renewcommand\]{\end{equation}}
\usepackage{verbatim}
\usepackage{xcolor}
\usepackage{amsmath, nccmath}
\usepackage{bm}
\usepackage{bbm}
\usepackage{lipsum}
\usepackage{stmaryrd}
\usepackage{wrapfig}
\usepackage{soul}
\begin{document}
\title{Mott versus hybridization gap in the low-temperature phase of $1T$-TaS$_2$}
\author{Francesco Petocchi}
\affiliation{Department of Physics, University of Fribourg, 1700 Fribourg, Switzerland}
\author{Christopher W. Nicholson}
\affiliation{Department of Physics, University of Fribourg, 1700 Fribourg, Switzerland}
\affiliation{Fritz-Haber-Institute der Max-Planck-Gesellschaft, Faradayweg 4-6, 14195 Berlin, Germany} 
\author{Bjoern Salzmann}
\affiliation{Department of Physics, University of Fribourg, 1700 Fribourg, Switzerland}
\author{Diego Pasquier}
\affiliation{Institute of Physics, Ecole Polytechnique F\'ed\'erale de Lausanne (EPFL), 1015 Lausanne, Switzerland}
\author{Oleg V. Yazyev}
\affiliation{Institute of Physics, Ecole Polytechnique F\'ed\'erale de Lausanne (EPFL), 1015 Lausanne, Switzerland}
\author{Claude Monney}
\affiliation{Department of Physics, University of Fribourg, 1700 Fribourg, Switzerland}
\author{Philipp Werner}
\affiliation{Department of Physics, University of Fribourg, 1700 Fribourg, Switzerland}
\begin{abstract}
We compute the correlated electronic structure of stacked $1T$-TaS$_2$ bilayers using the $GW$ + EDMFT method. Depending on the surface termination, the semi-infinite uncorrelated system is either band-insulating or exhibits a metallic surface state. For realistic values of the onsite and intersite interactions, a Mott gap opens in the surface state, but this gap is smaller than the gap originating from the bilayer structure. Our results are consistent with recent scanning tunneling spectroscopy measurements for different terminating layers, and with our own photoemission measurements, which indicate the coexistence of spatial regions with different gaps in the electronic spectrum.
\end{abstract}
\date{\today}
\maketitle
%
%
%
%
{\it Introduction}
The layered transition metal dichalcogenide 1$T$-TaS$_2$ has been studied intensively for decades, because it exhibits an intriguing interplay between lattice distortions and correlated electron phenomena. The material undergoes a series of charge density wave (CDW) transitions as temperature is lowered, first to an incommensurate CDW phase at 550~K, then to a nearly commensurate CDW phase around 350~K, and below 180~K to a commensurate CDW (CCDW) phase \cite{Wilson1975}. 
The in-plane periodic lattice distortion found in the CCDW state leads to the formation of star-of-David (SOD) clusters consisting of 13 Ta atoms, and the resistivity strongly increases. Since the monolayer system can be described by a half-filled Hubbard model on a triangular lattice, with each site representing a molecular orbital of the SOD cluster, 1$T$-TaS$_2$  in the CCDW state is often regarded as a polaronic Mott insulator \cite{Fazekas1979}. Interesting properties of this phase include a transition to a superconducting state under pressure \cite{Sipos2008}, and possible spin-liquid behavior \cite{Klanjsek2017}. 
It has also been shown that the CCDW phase can be switched into long-lived metallic metastable phases by the application of short laser or voltage pulses \cite{Stojchevska14,Cho2016}, which may be exploited in future memory devices \cite{Vaskivskyi2016}. The study of the equilibrium \cite{Cho2016,Butler2020,Lee2021,Wu2022}  and nonequilibrium \cite{Gerasimenko2019} phases of 1$T$-TaS$_2$ by scanning tunneling microscopy (STM) have furthermore revealed nontrivial patterns and spatial regions with different gaps in the electronic spectrum.

On the theory side, the electronic structure of the CCDW phase has been studied in several recent works \cite{Darancet2014,Ritschel2015,Ritschel2018,Lee2019,Shin2021,Lee2021}. Based on density functional theory (DFT) + $U$ calculations Ref.~\onlinecite{Darancet2014} argued that the system is Mott insulating in the in-plane direction, but metallic in the stacking (c-axis) direction, and suggested that the experimentally observed insulating nature of the CCDW phase may be due to stacking disorder, which has long been known to exist in this material \cite{Fung1980,Nakanishi1984}. This has led to further DFT investigations (supported by x-ray diffraction) into the role of the layer stacking and its effects on the electronic ground state \cite{Ritschel2015,Ritschel2018,Lee2019}.  Ref.~\onlinecite{Lee2019} showed that the lowest energy structure exhibits a specific stacking of bilayers, dubbed ``AL" stacking (with A referring to the center of the SOD and L to the upper right corner). 

While these DFT results suggest that hybridization gaps produced by the stacking arrangement of the 1$T$-TaS$_2$ monolayers make the low-temperature system insulating, recent STM measurements \cite{Butler2020,Lee2021} cannot be understood within a single-particle picture. In particular, Butler and coworkers \cite{Butler2020} observed different gaps in spectra obtained for surface terminations with a cleavage plane between two bilayers or within a bilayer. They argued that, in a weakly correlated material, a metallic state would be found for the single-layer termination, at odds with the experimental results. This calls for a systematic study of the correlated electronic structure in the stacked bilayer systems. 

In this work, we employ the $GW$ + extended dynamical mean field theory ($GW$+EDMFT) method \cite{Biermann2003,Ayral2013,Nilsson2017,Petocchi2021} to simulate semi-infinite systems of  1$T$-TaS$_2$ layers with the bilayer stacking identified in Refs.~\onlinecite{Lee2019,Butler2020,Lee2021} and for the two different surface terminations. In the absence of on-site and intersite interactions, our model calculations produce a metallic surface state for the termination within the bilayer, but this state undergoes a Mott transition if realistic interactions are added. Our results are in good agreement with previously published STM and our own spatially-resolved photoemission data, and clarify the interplay between hybridization gaps and Mott gaps in the CCDW phase of 1$T$-TaS$_2$.

\begin{figure}
\begin{minipage}{.55\linewidth}
\includegraphics[width=\textwidth]{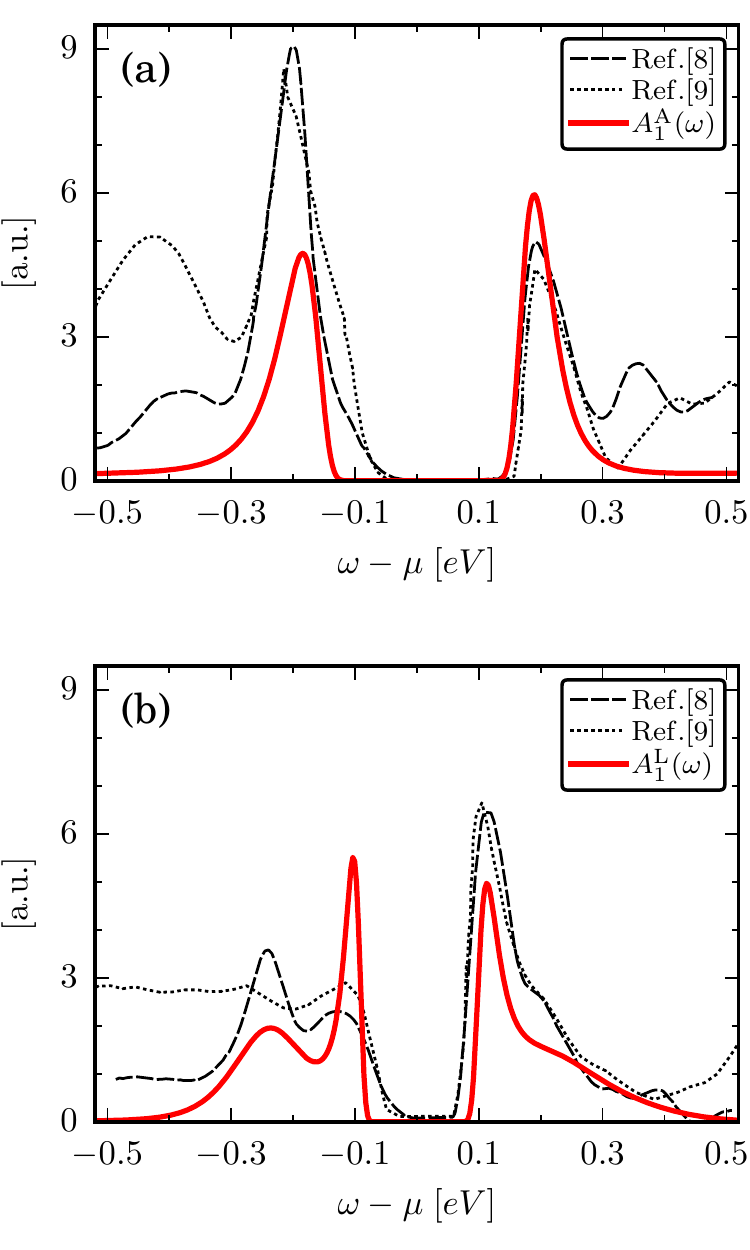}
\end{minipage} \hspace{0.4cm}
\begin{minipage}{.25\linewidth}
\includegraphics[width=\textwidth]{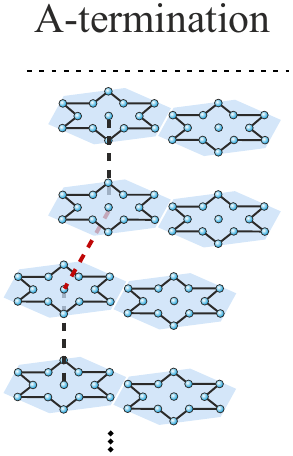}\vspace{0.8cm}
\includegraphics[width=\textwidth]{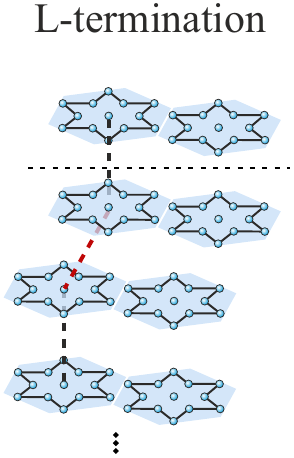}
\end{minipage}
\caption{Comparison between the $\mathbf{k}_{\sslash}$-integrated spectral function (red line) of the surface layer and the recent STM measurements of Ref.~\onlinecite{Butler2020,Lee2021} (black lines) for the A (a) and L (b) termination. In the right column we sketch the cleavage planes which result in the two types of terminations. Black and red dashed lines indicate, respectively, the vertical hoppings $t_{\mathrm{ab}}^\mathrm{A}$ and $t_{\mathrm{ab}}^\mathrm{L}$ within the supercell.}
\label{Figure:STM}
\end{figure}

{\it Model and metod.}
The effective noninteracting single-band Hamiltonian $\mathcal{H}_{\mathrm{Ta}}$ for the SOD clusters of the $1T$-TaS$_2$ monolayer was obtained with density functional theory (DFT) as described in Ref.~\onlinecite{pasquier2021textitab} and the Supplemental Material (SM). Our goal here is to model a system consisting of a semi-infinite sequence of bilayers with AL stacking and the two different surface terminations, as illustrated by the sketches in Fig.~\ref{Figure:STM}. For this, we explicitly construct a Hamiltonian $\mathcal{H}_{\mathrm{ab}}$ for eight layers, labeled by roman letters $\mathrm{a,b}=1,\ldots,8$, and periodically repeat the solution for layers 7 and 8 to mimic an infinite bulk. The AL stacking configuration features bilayers in which the SOD centers are aligned in the c-direction, whereas a shift of $-2\mathbf{a}$ is present between two neighboring bilayers (stacking vector $\mathbf{T}_{S}=-2\mathbf{a}+2\mathbf{c}$, with $\mathbf{a},\mathbf{b},\mathbf{c}$ denoting the primitive vectors of the unit cell). We will denote the setup with cleavage plane between bilayers as ``A termination", and the setup with the cleavage plane within a bilayer as ``L termination". Several {\it ab-inito} studies emphasized that the overlap between the $d_{z^2}$-like effective orbitals results in strong inter-plane hopping within a bilayer, while the hybridizations between bilayers are weaker. 
In order to build $\mathcal{H}_{\mathrm{ab}}$ we introduce two off-diagonal hopping parameters $t_{\mathrm{ab}}^{\mathrm{A}}$ and  $t_{\mathrm{ab}}^{\mathrm{L}}$ connecting, respectively, nearest-neighbor layers forming a bilayer and shifted layers:
\begin{equation}
\mathcal{H}_{\mathrm{ab}}(\mathbf{R}_{i},\mathbf{R}_{j})=\mathcal{H}_{\mathrm{Ta}}(\mathbf{R}_{i},\mathbf{R}_{j})\delta_{\mathrm{ab}}-t_{\mathrm{ab}}^{\mathrm{A}}\delta_{\mathbf{R}_{i},\mathbf{R}_{j}}-t_{\mathrm{ab}}^{\mathrm{L}}(\mathbf{R}_{i},\mathbf{R}_{j}),
\end{equation}
where $ \mathbf{R}_{i}\equiv \left\{ \mathbf{R}_{x},\mathbf{R}_{y}\right\}_i$ are the vectors of the unit cell. 
The Fourier transform of $t_{\mathrm{ab}}^{\mathrm{A}}$ has no component in the $\left\{ \mathbf{k}_{x},\mathbf{k}_{y}\right\} \equiv \mathbf{k}_{\sslash}$ plane. After a $-2\mathbf{a}$ translation, a cluster orbital has three neighbors in the adjacent layer, one in the same unit cell and two in neighboring unit cells, leading to an off-diagonal or $\mathbf{k}_{\sslash}$-dependent hybridization. While the distances to the different unit cells are not identical, given their small difference, we connect the three unit cells with the same $t_{\mathrm{ab}}^{\mathrm{L}}$.
The $GW$+EDMFT method \cite{Ayral2013,Werner2016,Boehnke2016,Ayral2017,Nilsson2017,Petocchi2019,Petocchi2020,Petocchi2021,chen2021causal} employed to study the effect of electronic correlations is similar to the real-space extension which has recently been applied to compounds with several sites in the unit cell \cite{Petocchi2019,Petocchi2020,Petocchi2021} (see SM).  To each layer, we associate an EDMFT-type impurity problem with fermionic and bosonic Weiss fields, which is solved using a continuous-time Monte Carlo method capable of dealing with retarded interactions \cite{Werner2006,Werner2010,Hafermann2013}. These solutions provide a set of local self-energies $\Sigma^{\mathrm{EDMFT}}_{\mathrm{aa}}\left(i\omega_n\right)$ and polarizations $\Pi^{\mathrm{EDMFT}}_{\mathrm{aaaa}}(i\Omega_n)$ which  replace the corresponding components of the local projection of their $GW$ counterparts (calculated for an 8-atom supercell and $20 \times 20$ $\mathbf{k}_{\sslash}$-points). The result is a momentum-dependent self-energy $\Sigma_{\mathrm{ab}}(\mathbf{k}_{\sslash},i\omega_n)$ and polarization $\Pi_{\mathrm{adbc}}(\mathbf{k}_{\sslash},i\Omega_n)$ which incorporates the effects of strong local and weaker nonlocal interactions. In the last step of the $GW$+EDMFT self-consistency loop $\Sigma_{\mathrm{ab}}(\mathbf{k}_{\sslash},i\omega_n)$ and $\Pi_{\mathrm{adbc}}(\mathbf{k}_{\sslash},i\Omega_n)$ are used to compute the lattice Green's function $G_{\mathrm{ab}}(\mathbf{k}_{\sslash},i\omega_n)$ and screened interaction $W_{\mathrm{adbc}}(\mathbf{k}_{\sslash},i\Omega_n)$. Given the insulating nature of the material, we 
assume a Coulomb-like bare interaction which depends on the spatial coordinates as 
\begin{equation}
U_{\mathrm{aabb}}(\mathbf{R}_{i},\mathbf{R}_{j})=U\delta_\mathrm{ab}\delta_{\mathbf{R}_{i},\mathbf{R}_{j}}+\frac{V}{\left|\mathbf{R}_{i}-\mathbf{R}_{j}+\mathbf{r}_{\mathrm{a}}-\mathbf{r}_{\mathrm{b}}\right|},
\end{equation}
where $U$ is the local Hubbard interaction, $V$ a parameter that sets the strength of the non-local density-density interactions and $\mathbf{r}_{\mathrm{a}}$ the position of the cluster orbital along the c-axis. With $\mathcal{H}_{\mathrm{ab}}(\mathbf{k}_{\sslash})$ and $U_{\mathrm{aabb}}(\mathbf{k}_{\sslash})$ as input, the $GW$+EDMFT scheme provides a self-consistent solution where the effective local interaction $\mathcal{U}(\omega)$ on a given cluster site is screened by non-local charge fluctuations. 

To capture the effect of the semi-infinite bulk, we add an embedding potential $E(\mathbf{k}_{\sslash},i\omega_n)$ to the eighth layer. $E$ is computed with a continued fraction recursive formula that periodizes the properties of the last two layers:
\begin{equation}
E(\mathbf{k}_{\sslash},i\omega_{n})=\frac{t_{\mathrm{67}}^{2}}{z_{7}-\frac{t_{\mathrm{78}}^{2}}{z_{8}-\frac{t_{\mathrm{67}}^{2}}{z_{7}-\ldots}}} \, ,
\end{equation}
where $t_{\mathrm{ab}}^{2}=\mathcal{H}_{\mathrm{ab}}(\mathbf{k}_{\sslash})\mathcal{H}_{\mathrm{ba}}^{*}(\mathbf{k}_{\sslash})$, with $\mathrm{ab}\in\left\{ 67,78\right\} $, and $z_{\mathrm{a}}=i\omega_{n}+\mu-\mathcal{H}_{\mathrm{Ta}}(\mathbf{k}_{\sslash})-\Sigma_{\mathrm{aa}}(\mathbf{k}_{\sslash},i\omega_{n})$ with $\mathrm{a}\in\left\{ 7,8\right\}$. As we will see below, eight layers are enough to reach the bulk behavior, where the self-energy becomes layer-independent. All calculations are performed at half-filling, i.e., with eight electrons in the supercell, and at a temperature of 30~K.

{\it Results.}
Our model of the layered structure depends on four parameters: the hoppings in the vertical direction $t_{\mathrm{ab}}^{\mathrm{A}}$ and $t_{\mathrm{ab}}^{\mathrm{L}}$, the local Hubbard interaction $U$ and the magnitude of the nonlocal density-density interaction $V$. To determine these parameters we first applied $GW$+EDMFT to the $1T$-TaS$_2$ monolayer for which STM measurements are available and show a gap $\Delta \sim 0.4$~eV \cite{Lin2020}. A good agreement with the STM spectra of Ref.~\onlinecite{Lin2020} is obtained for $U=0.4$~eV and $V=0.08$~eV \cite{chen2021causal}.
The values of $t_{\mathrm{ab}}^{\mathrm{A,L}}$ were subsequently determined by reproducing the STM measurements reported in two recent studies \cite{Butler2020,Lee2021} on the surface effects of stacking ordering in $1T$-TaS$_2$. This yields $t_{\mathrm{ab}}^{\mathrm{A}}$=0.2~eV and $t_{\mathrm{ab}}^{\mathrm{L}}$=0.045~eV. The above set of Hamiltonian parameters allows to reproduce the main features of the experimental spectra, as shown in Fig.~\ref{Figure:STM}, which plots in black the recent STM measurements and in red the theoretical $\mathbf{k}_{\sslash}$-integrated surface spectral function $A_{1}^{\mathrm{A,L}}(\omega)$. In particular, for both setups one obtains a good match for the gap size, with $\Delta_{1}^{\mathrm{A}}\sim0.4$~eV ($\Delta_{1}^{\mathrm{L}}\sim0.22$ eV) in the system with A (L) termination. In the case of L termination, also the higher-energy spectral features look consistent with the experimental data. STM features beyond $\pm 0.4$~eV most likely originate from bands which are not contained in our low energy model.

\begin{figure}
\begin{center}
\includegraphics[width=0.48\textwidth]{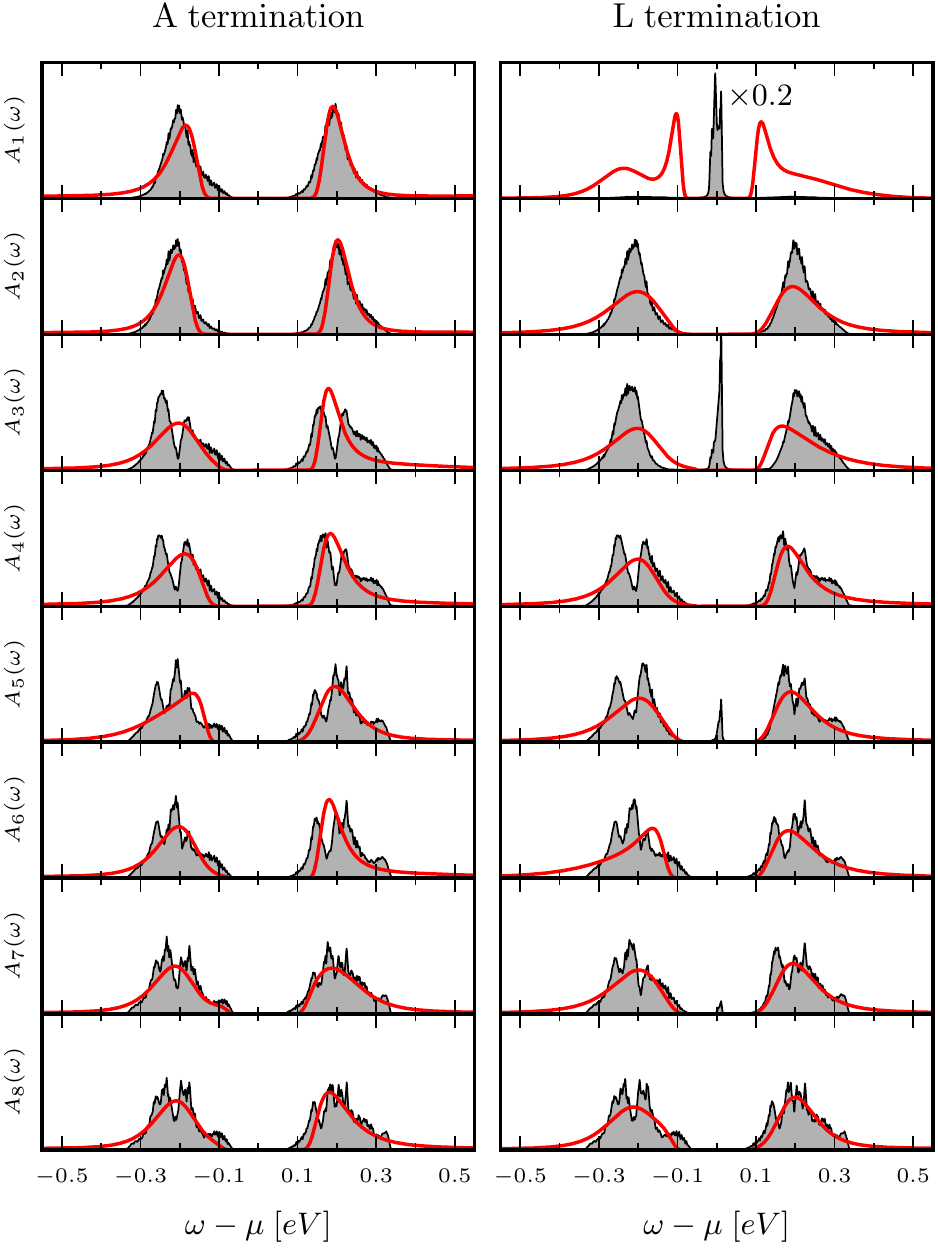}
\caption{Layer resolved $\mathbf{k}_{\sslash}$-integrated spectral function for the setup with A termination (left) and L termination (right). The $GW$+EDMFT results (red lines)  are superimposed to grey regions indicating the spectral functions of the non-interacting model. Without interactions, the model with $\mathrm{L}$ termination hosts a metallic state pinned to the surface and decaying inside the structure. This surface state undergoes a Mott transition if the interactions are included. The non-interacting spectrum of the surface layer with L termination has been scaled with a factor 0.2 for graphical purposes. The appearance of the metallic peaks on odd layers reflects the two-atom unit cell. }
\label{Figure:Hetero}
\end{center}
\end{figure}

The layer-resolved spectral functions are reported in Fig.~\ref{Figure:Hetero}, where one notices that, already at the non-interacting level (grey regions), our model produces qualitatively different results for the two terminations. 
In the case of the A termination, where no bilayers are broken, the system is a band insulator, which suggests that even in the presence of sizeable interactions, the insulating character is primarily due to bonding/antibonding splittings. In contrast, the uncorrelated setup with L termination hosts a metallic state at the surface, which extends only a few layers into the bulk. It is worth noticing that the embedding potential prevents the appearance of a surface state at the bottom of the eight-layer structure. The metallic peak in the layer-resolved densities of states disappears everywhere when the interactions are included, with a clear splitting of the peak of the surface layer into lower and upper Hubbard bands. This shows that the insulating nature of the  system with L termination is, in the surface region, the result of Mott physics. In the SM we provide further evidence of the correlation-driven insulating state in the surface layer by plotting the imaginary part of the EDMFT self-energies. Im$\Sigma^{\mathrm{EDMFT}}_{\mathrm{aa}}(i\omega_n)$ is very small and vanishes for $\omega_n\rightarrow0$ in all the layers except for the surface layer with L termination, where the low-frequency behavior shows a divergence as one expects for a paramagnetic Mott insulator. Hence, the insulating nature of the system with L termination results from a combination of band-insulating and Mott insulating behavior. While the bulk of perfectly stacked $1T$-TaS$_2$ is a band insulator with a hybridization gap induced by the strong $t^\mathrm{A}$ hopping within the bilayers, the top layer of the system with L termination, or an un-paired layer within a bulk with stacking disorder, behaves like a Mott insulating mono-layer of $1T$-TaS$_2$.
\begin{figure}
\begin{center}
\includegraphics[width=0.48\textwidth]{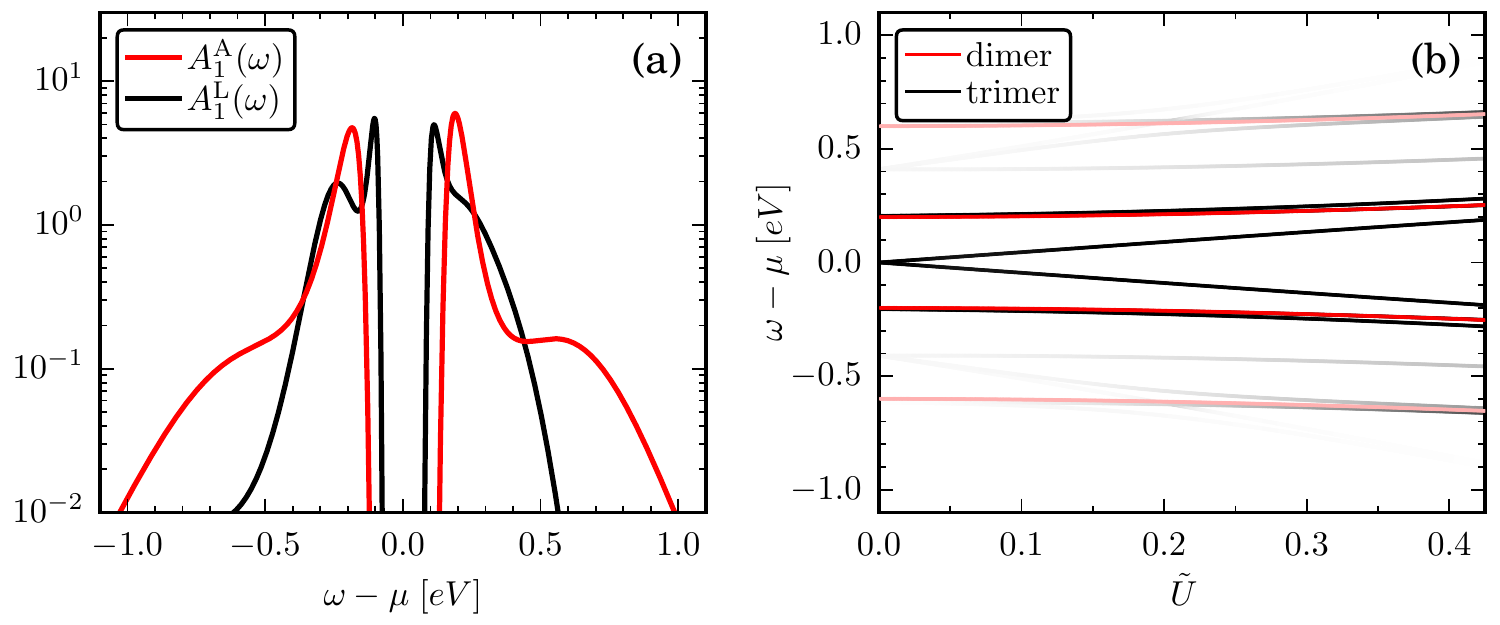}\vspace{0.4cm}
\includegraphics[width=0.48\textwidth]{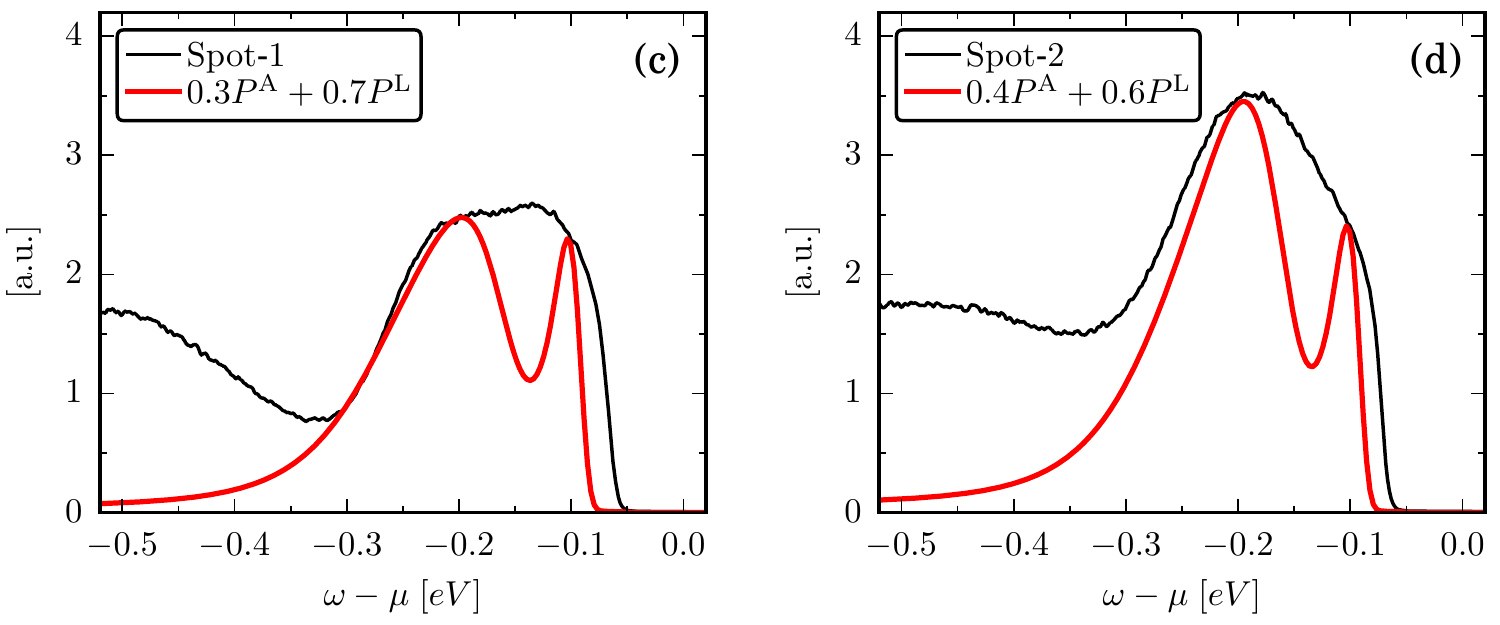}
\caption{The log-plot in panel (a) magnifies the high energy features of the $\mathbf{k}_{\sslash}$-integrated spectral function of the surface layer for the two terminations. Red (black) lines in panel (b) indicate the excitation energies of the Hubbard dimer (trimer) as a function of the local interaction $\tilde U$, where the color intensity indicates the $\tilde U$-dependent weight of the different excitations. Black lines in panels (c) and (d) show the experimental PES profiles obtained in different regions of a $1T$-TaS$_2$ sample, while red lines indicate the theoretical PES signals averaged over the two terminations as described in the text. The peak widths in the theoretical spectra are controlled by the parameters of the analytical continuation procedure.}
\label{Figure:PES_Model}
\end{center}
\end{figure}
\begin{figure*}
\includegraphics[width=0.49\textwidth]{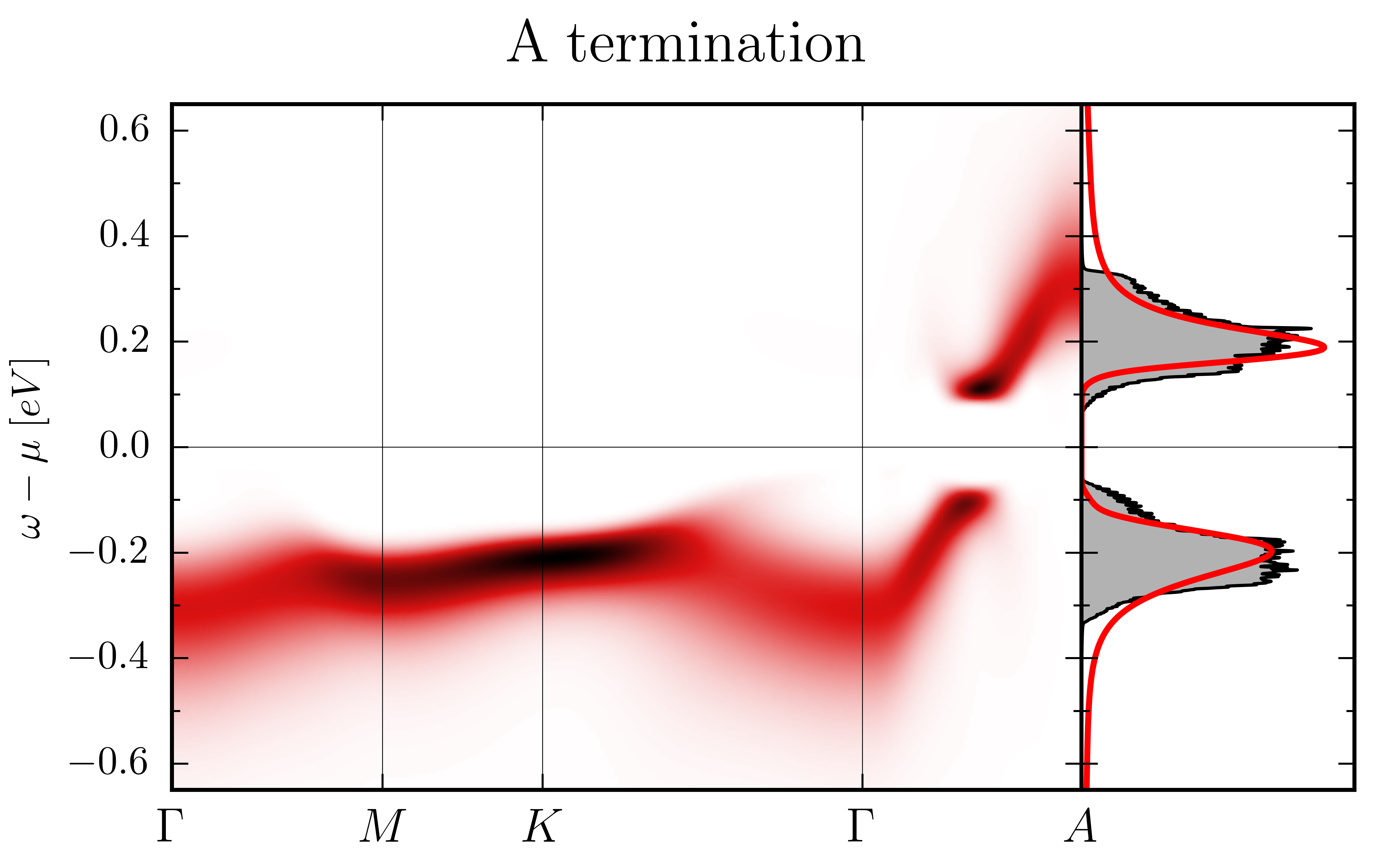}
\includegraphics[width=0.49\textwidth]{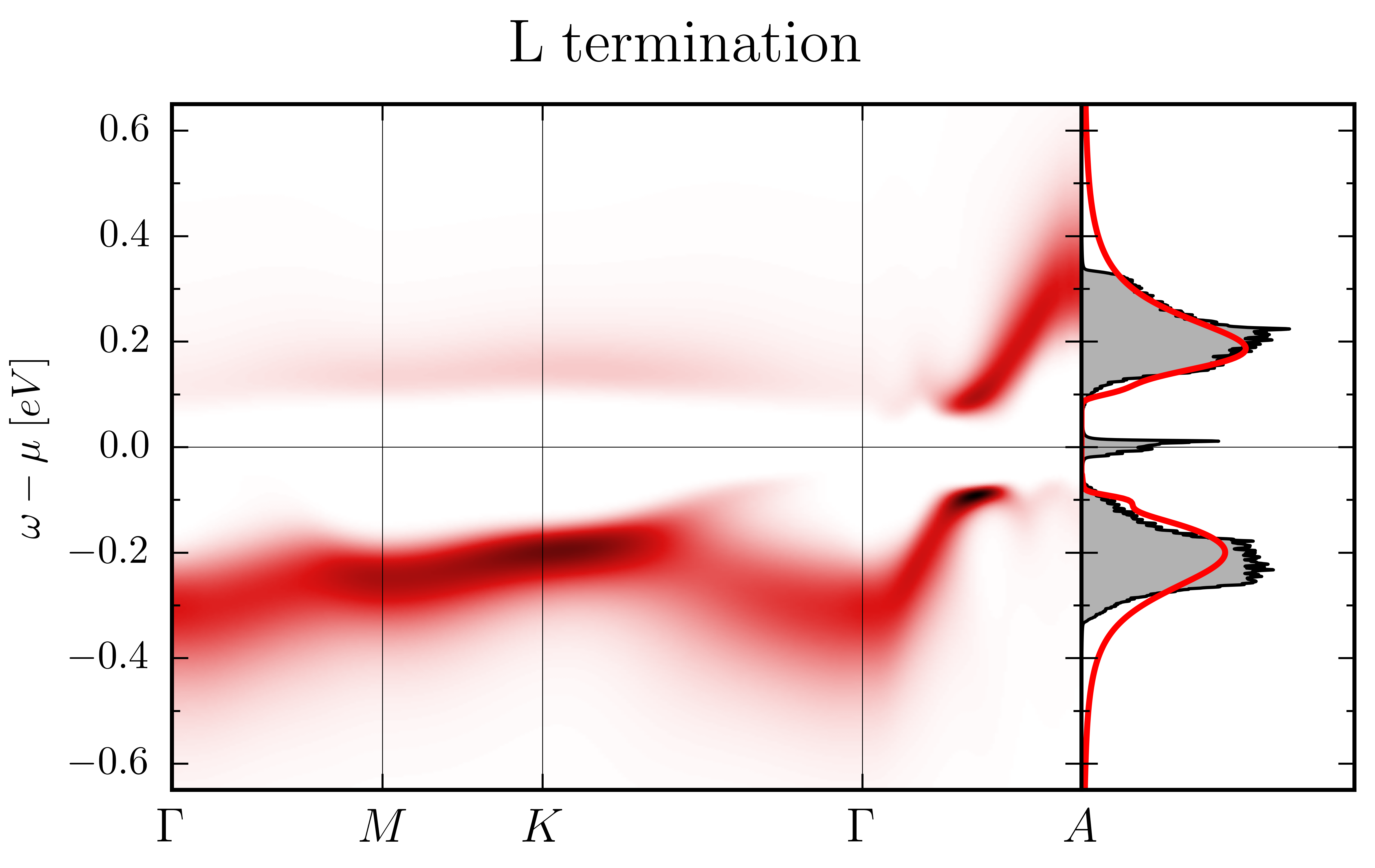}
\caption{Momentum-resolved spectral functions of the interacting system for the two types of terminations. The dispersion along the $\Gamma$-$A$ direction has been obtained by Fourier transforming the layer-resolved spectral function at each $\mathbf{k}$ point along the $\Gamma$-$M$-$K$-$\Gamma$ path. Also indicated on the right are the local spectral functions for the non-interacting (gray) and interacting (red) model. }
\label{Figure:Akw}
\end{figure*}

To clarify the nature of the peaks in the surface-layer spectra $A_{1}^{\mathrm{A,L}}(\omega)$ we solved a two-site Hubbard model (dimer) with an inter-site hopping $t=0.2$ eV and a three-site Hubbard model (trimer) with $t_{12}=0.045$ eV and $t_{23}=0.2$ eV. To label the Hamiltonian blocks we considered only the charge quantum number and computed, as a function of the on-site repulsion $\tilde U$, the ground states with $N=2$ and $3$, respectively, plus the eigenenergies of the two adjacent charge sectors $N\pm 1$. The single-particle excitation energies with respect to the Fermi level $\pm\left(E_{n}^{N\pm1}-E_{0}^{N}\right)$ of the dimer (trimer) are shown by the red (black) lines in Fig.~\ref{Figure:PES_Model}(b). The line width is indicative of the relative weight of the pole in the local spectrum $-\frac{1}{\pi}\text{Im}G_{\mathrm{a}}(\omega)$ (in the trimer case we considered the weakly hybridized site). In the dimer results, representative of the system with A termination, one notices that the spectrum is gapped even at $\tilde U=0$, while an increase of the local interaction splits the poles further. These observations are in agreement with the non-interacting spectra of Fig.~\ref{Figure:Hetero}. The red poles closest to the Fermi level correspond to electron and hole excitations to bonding/antibonding states, $\left|\psi_{0}^{N=3}\right\rangle =\frac{1}{\sqrt{2}}\left(\left|\uparrow\downarrow,s\right\rangle -\left|s,\uparrow\downarrow\right\rangle \right)$ and $\left|\psi_{0}^{N=1}\right\rangle =\frac{1}{\sqrt{2}}\left(\left|0,s\right\rangle +\left|s,0\right\rangle \right)$, where $s$=$\uparrow$ and $s$=$\downarrow$ are degenerate. The trimer model has an additional pole at zero frequency, which is split by the interaction into two excitation energies that, for relevant values of $\tilde U$, remain separated from the bonding-antibonding states at higher energies. While the $\tilde U$ parameters of the simple dimer and trimer models cannot be directly compared with the bare $U$ or effective $\mathcal{U}(\omega)$ interactions in the full calculation, for $\tilde U$ lower than the bare $GW$+EDMFT interaction of $U=0.4$~eV the dimer exhibits a larger gap than the trimer, which is consistent with the surface spectral functions reported on a log-scale in Fig.~\ref{Figure:PES_Model}(a), and with our interpretation in terms of a bonding/antibonding and Mott gap. Furthermore, the high-energy poles of the two minimal models consistently explain the high-energy spectral weight obtained in the full calculations. 

In agreement with several experimental observations \cite{Perfetti2006,PLigges2018,Perfetti2008,Avigo2019}, our spatially-resolved photoemission spectroscopy (PES) data, shown in Fig.~\ref{Figure:PES_Model}, indicate the presence of two substructures with a separation of about 0.1~eV close to the Fermi level. By performing a raster scan of the sample surface with a focused 6.2 eV laser source, we found that these substructures vary in relative intensity across the sample (see SM). Based on the different gaps obtained for the A and L termination, and the short penetration depth of photoemission, we argue that this observation is due to the superposition of PES signals from different spatial domains with the two types of terminations. Assuming that electrons contributing to the PES signal are emitted with a probability that decays exponentially from the surface of the sample, proportional to $\exp(-|z|/|\mathbf{c}|)$, the theoretical PES for the two terminations $P^{\mathrm{A,L}}$ can be computed as a weighted average of the spectra over the layers, multiplied by the Fermi distribution for 30~K. The red lines in Fig.~\ref{Figure:PES_Model}(c,d) show that different weighted averages allow to qualitatively explain the two substructure seen in the experimental spectra.

The insulating solution that we find for all layers when correlations are included implies that the electronic bands of the system are gapped also along the c-axis. Even though we do not have a periodic structure in the c-direction, we can nevertheless provide indications on how the dispersion along $\Gamma$-$A$ looks like. We first interpolate the layer-resolved Green's function along a high-symmetry path within the planar Brillouin zone and then compute the Fourier transform along the $z$ direction as
\begin{equation}
G(\mathbf{k}_{\sslash},\mathbf{k}_{z},\tau)=\frac{1}{8}\underset{\mathrm{ab}}{\sum}e^{i\mathbf{k}_{z}\left(\mathbf{z}_{\mathrm{a}}-\mathbf{z}_{\mathrm{b}}\right)}G_{\mathrm{ab}}(\mathbf{k}_{\sslash},\tau),
\end{equation}
where $\mathbf{k}_{\sslash}\in\left\{ \Gamma-M-K-\Gamma\right\}$ and $\mathbf{k}_{z}\in\left\{ \Gamma-A\right\}$. The maximum entropy method \cite{Jarrell1996} is then used for the analytical continuation to the real-frequency axis. The result is shown in Fig.~\ref{Figure:Akw}, while the momentum-resolved spectral functions for the different layers are reported in the SM. As expected, for both terminations we find a similar gap of the spectrum along the $\Gamma$-$A$ direction. The most notable difference is a weak and weakly dispersive band in the unoccupied part of the spectrum for the L termination, the upper Hubbard band of the surface layer, which may be detectable with two-photon photoemission.

{\it Conclusions.} We solved a minimal multi-layer model to clarify how electronic correlations affect the bulk and surface states of $1T$-TaS$_2$ in the low-temperature CCDW phase. In particular, we demonstrated the importance of Mott physics in gapping out a surface state in the system with L termination, while the bulk layers are essentially band insulating. For appropriate inter-plane hopping amplitudes, our results are in remarkably good agreement with recent STM measurements showing different gap sizes depending on the cleavage plane, and they provide a natural explanation of the presented photoemission data in terms of a superposition of different spatial regions with A and L termination. In both cases we can interpret the high-energy substructures of our model spectra (near $\pm0.4$ eV) as originating from a hybridization gap, while the lower-energy peaks can be associated with the Mott insulating surface state. Our results provide a solid basis for the previous interpretations of the STM measurements \cite{Butler2020,Lee2021} and show that the surface region of 1$T$-TaS$_2$ in the CCDW phase exhibits a nontrivial interplay between band insulating and Mott insulating behavior.

{\it Acknowledgments}
F.P., D.P., O.V.Y. and P.W. acknowledge support from the Swiss National Science Foundation through NCCR MARVEL. F.P. and P.W. acknowledge support from the European Research Council through ERC Consolidator Grant 724103. C.W.N., B.S. and C.M. acknowledge the support from the Swiss National Science Foundation Grant No. P00P2\textunderscore170597. The calculations were performed on the Beo05 clusters at the University of Fribourg and the Piz Daint cluster at the Swiss National Supercomputing Centre (CSCS) under projects ID mr26 and s1008. 
\bibliography{paper}

\onecolumngrid

\newpage
\section*{Supplemental Material: Mott versus hybridization gap in the low-temperature phase of $1T$-TaS$_2$}

\subsection*{Calculation of the Wannier Hamiltonian $\mathcal{H}_{\mathrm{Ta}}$ }
To obtain the noninteracting one-band Hamiltonian $\mathcal{H}_{\mathrm{Ta}}$ for the TaS$_2$ monolayer in the commensurate charge-density-wave (CCDW) phase, we used a combination of density functional theory (DFT) and wannierization approaches.

DFT calculactions were carried out using the \textsc{Quantum} ESPRESSO package \cite{giannozzi2009quantum}. The exchange-correlation functional was approximated by the generalized-gradient parametrization according to Perdew, Burke and Ernzerhof \cite{perdew1996generalized}. The interactions between core and valence electrons were described using projector-augmented-wave pseudopotentials, including the $s$ and $p$ semicore states of the Ta atoms explicitly. The wave-function and charge-density cutoffs were set to $60$ and $300$ Ry. We have used a grid of $8 \times 8$ k-points and a Marzari-Vanderbilt smearing of $0.01$ Ry.

The structure of the CCDW phase was calculated by relaxing the atomic positions and lattice parameters in a $\sqrt{13}\times \sqrt{13}$ supercell containing thirteen Ta atoms.
\begin{figure}[h]
    \centering
    \includegraphics[height=8cm]{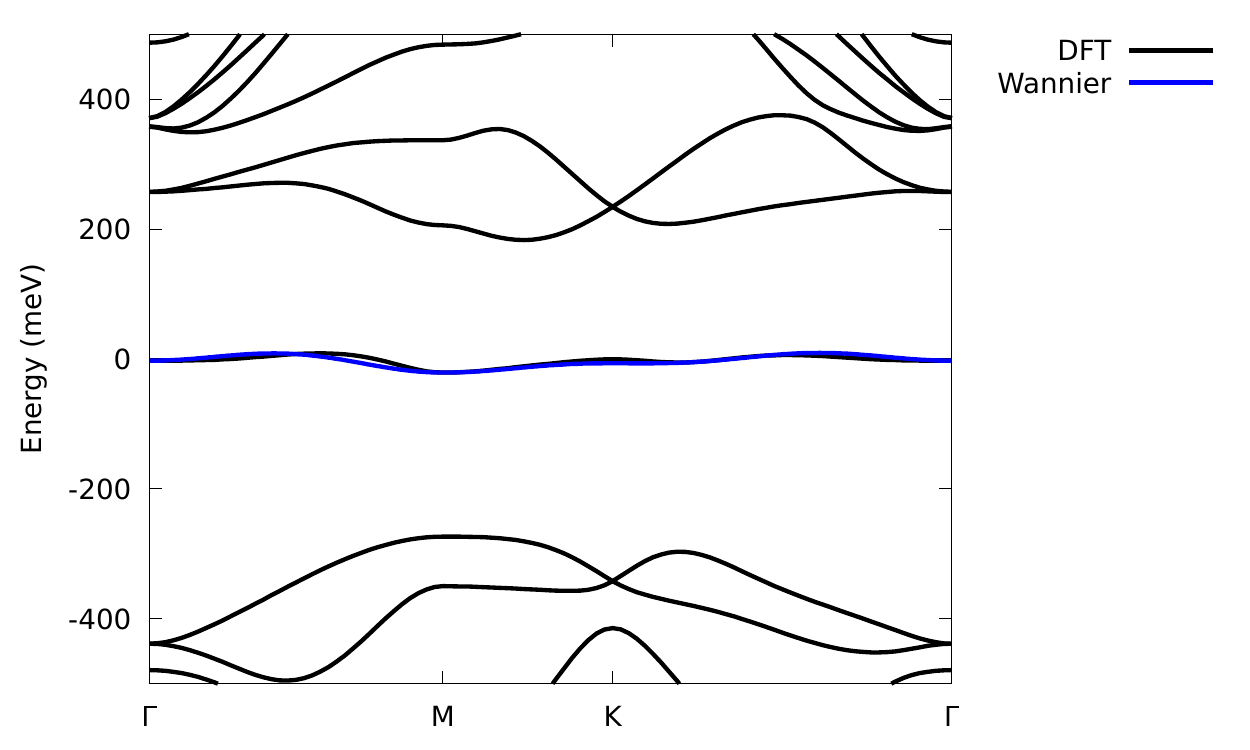}
    \caption{\label{fig:wannier} DFT and Wannier-interpolated band structure for the TaS$_2$ monolayer in the CCDW phase. The Fermi energy is set to zero.}
    \label{Fig:AkwLres}
\end{figure}
The wannierization was carried out using the Wannier90 code \cite{mostofi2008wannier90}. For the Wannier transformation, we only included the half-filled narrow band that crosses the Fermi level. Since this band is separated in energy from the other ones, the transfomation is uniquely defined and there is no need to specify a disentanglement or a gauge-fixing prescription. The Wannier function was computed on a grid of $4\times4$ k-points.

Figure~\ref{fig:wannier} shows the DFT band structure for a TaS$_2$ monolayer in the CCDW phase, as well as the one obtained with the interpolated Wannier Hamiltonian.

\newpage

\subsection*{$GW$+EDMFT method}
The GW+EDMFT method is a numerical scheme that incorporates in the same self-consistency loop the effects of charge fluctuations and strong (weak) local (nonlocal) interactions.\cite{Biermann2003,Ayral2013,Nilsson2017,Petocchi2021} The nonlocal contributions to the self-energy $\Sigma$ and polarization $\Pi$ are computed at the $GW$ level and the corresponding local projections are replaced by the self-energy and polarization provided by extended DMFT (EDMFT).\cite{Sun2002} This procedure avoids a double counting of self-energy and polarization diagrams. The resulting $\Sigma$ and $\Pi$ are then used to compute the lattice Green's function $G$ and screened interaction $W$. Finally, two coupled self-consistency equations yield the fermionic $\mathcal{G}$ and bosonic $\mathcal{U}$ ``Weiss fields" of the impurity model. 

Fully {\it ab-inito} implementations incorporate the effects of bands residing outside the low-energy model, namely Tier-III, by means of an additional single-shot $G^0W^0$ self-energy and a frequency-dependent bare interaction screened by high energy processes.\cite{Nilsson2017,Petocchi2019,Petocchi2020,Petocchi2021} The implementation presented in the manuscript is not fully  {\it ab-inito}, as it lacks these $G^0W^0$ self-energy contribution and considers a static, but momentum dependent, bare interaction. However, it goes beyond a simple model description as it starts from a DFT-derived bandstructure and employs a non-local interaction that takes into account the spatial arrangements of molecular orbitals within the multi-layer system. 

Our self-consistency loop is a real-space extension of the $GW$+EDMFT method,\cite{Petocchi2020,Petocchi2021} where a single-orbital impurity problem is solved for each layer of the eight-layer structure. Starting from some initial guess for the eight impurity self-energies $\Sigma_{\mathrm{aa}}^{\mathrm{EDMFT}}$ and polaritazions $\Pi_{\mathrm{aaaa}}^{\mathrm{EDMFT}}$ the algorithm performs the following steps:
\begin{enumerate}
\item The momentum-dependent self-energy $\Sigma_{\mathbf{k}}^{GW}$ and polarization $\Pi_{\mathbf{q}}^{GG}$ in the $GW$ approximation are computed as a function of the planar momentum for a supercell containing eight sites in the unit cell $\mathrm{a,b,c,d}=1,\ldots,8$:
\begin{itemize}
\item $\Pi^{GG}_{\mathrm{adbc}}\left(\mathbf{k}_{\sslash},\tau\right)=\underset{\mathbf{q}_{\sslash}}{\sum}G_{\mathrm{ab}}\left(\mathbf{k}_{\sslash},\tau\right)G_{\mathrm{cd}}\left(\mathbf{q}_{\sslash}-\mathbf{k}_{\sslash},-\tau\right)$,
\item $\Sigma^{GW}_{\mathrm{ab}}\left(\mathbf{k}_{\sslash},\tau\right)=-\underset{\mathbf{q}_{\sslash}}{\sum}G_{\mathrm{dc}}\left(\mathbf{k}_{\sslash},\tau\right)W_{\mathrm{adbc}}\left(\mathbf{q}_{\sslash}-\mathbf{k}_{\sslash},\tau\right)$.
\end{itemize}
The local projections on each site are then replaced with the EDMFT counterparts:
\begin{itemize}
\item \ensuremath{\Pi_{\mathrm{adbc}}\left(\mathbf{k}_{\sslash},i\Omega_{n}\right)=\Pi_{\mathrm{adbc}}^{GG}\left(\mathbf{k}_{\sslash},i\Omega_{n}\right)-\Pi_{\mathrm{aaaa}}^{GG}\left(i\Omega_{n}\right)|_{\mathrm{loc}}+\Pi_{\mathrm{aaaa}}^{\mathrm{EDMFT}}\left(i\Omega_{n}\right)},
\item \ensuremath{\Sigma_{\mathrm{ab}}\left(\mathbf{k}_{\sslash},i\omega_{n}\right)=\Sigma_{\mathrm{ab}}^{GW}\left(\mathbf{k}_{\sslash},i\omega_{n}\right)-\Sigma_{\mathrm{aa}}^{GW}\left(i\omega_{n}\right)|_{\mathrm{loc}}+\Sigma_{\mathrm{aa}}^{\mathrm{EDMFT}}\left(i\omega_{n}\right)}.
\end{itemize}
where the roman indices refer to the different layers. It is important to note that, even if strong correlations are treated locally, i.e. within a given layer, the $GW$ calculation contributes inter-layer terms to $\Sigma$ and non-density-density components to $\Pi$.
\item Use the polarization and self-energy, as well as the bare interaction $U_{\mathrm{aabb}}\left(\mathbf{k}_{\sslash}\right)$ to compute the local screened interaction and local lattice Green's function:
\begin{itemize}
\item $W_{\mathrm{adbc}}\left(i\Omega_{n}\right)=\sum_{\mathbf{k}_{\sslash}}U_{\mathrm{aabb}}\left(\mathbf{k}_{\sslash}\right)\left[1-\Pi_{\mathrm{adbc}}\left(\mathbf{k}_{\sslash},i\Omega_{n}\right)U_{\mathrm{aabb}}\left(\mathbf{k}_{\sslash}\right)\right]^{-1}$,
\item $G_{\mathrm{ab}}\left(i\omega_{n}\right)=\sum_{\mathbf{k}_{\sslash}}\left[\left(i\omega_{n}+\mu\right)1-\ensuremath{\mathcal{H}_{\mathrm{ab}}\left(\mathbf{k}_{\sslash}\right)}-\Sigma_{\mathrm{ab}}\left(\mathbf{k}_{\sslash},i\omega_{n}\right)\right]^{-1}$.
\end{itemize}
\item Impose the two self-consistency conditions for each layer:
\begin{itemize}
\item $W_{\mathrm{aaaa}}=W^{\mathrm{imp}}_{\mathrm{a}}$,
\item $G_{\mathrm{aa}}=G^{\mathrm{imp}}_{\mathrm{a}}$.
\end{itemize}
\item For every layer, compute the bosonic and fermionic Weiss fields $\mathcal{U}$ and $\mathcal{G}$ of the EDMFT impurity problems:
\begin{itemize}
\item $\mathcal{U}_{\mathrm{a}}=W_{\mathrm{aaaa}}\left[1+\Pi_{\mathrm{aaaa}}^{\mathrm{EDMFT}}W_{\mathrm{aaaa}}\right]^{-1}$,
\item $\mathcal{G}_{\mathrm{a}}=\left[G_{\mathrm{aa}}^{-1}-\Sigma_{\mathrm{aa}}^{\mathrm{EDMFT}}\right]^{-1}$.
\end{itemize}
\item The solution of the eight impurity problems is obtained with a continuous-time Monte Carlo solver for models with dynamically screened interactions,\cite{Werner2006,Werner2010,Hafermann2013} which provides eight pairs of density-density correlation functions $\chi^{\mathrm{imp}}$ and impurity Green's functions $G^{\mathrm{imp}}$:
\begin{itemize}
\item $\chi^{\mathrm{imp}}_{\mathrm{a}}=\left\langle \hat{n}_{\mathrm{a}}\left(\tau\right)\hat{n}_{\mathrm{a}}\left(0\right)\right\rangle $,
\item $G^{\mathrm{imp}}_{\mathrm{a}}$,
\end{itemize}
which are used to solve two Dyson equations, and to extract the local EDMFT polarization and self-energy:
\begin{itemize}
\item $\Pi_{\mathrm{aaaa}}^{\mathrm{EDMFT}}=\chi^{\mathrm{imp}}_{\mathrm{a}}\left[\mathcal{U}_{\mathrm{a}}\chi^{\mathrm{imp}}_{\mathrm{a}}-1\right]^{-1}$,
\item $\Sigma_{\mathrm{aa}}^{\mathrm{EDMFT}}=\mathcal{G}_{\mathrm{a}}^{-1}-\left(G^{\mathrm{imp}}_{\mathrm{a}}\right)^{-1}$.
\end{itemize}
The site-diagonal $\Pi^{\mathrm{EDMFT}}$ and $\Sigma^{\mathrm{EDMFT}}$ are then substituted back into the first step, and the loop is iterated until a converged solution is obtained. 
\end{enumerate}

\newpage

\subsection*{Layer-resolved self-energy}
\begin{figure}[h]
    \centering
    \includegraphics[height=14cm]{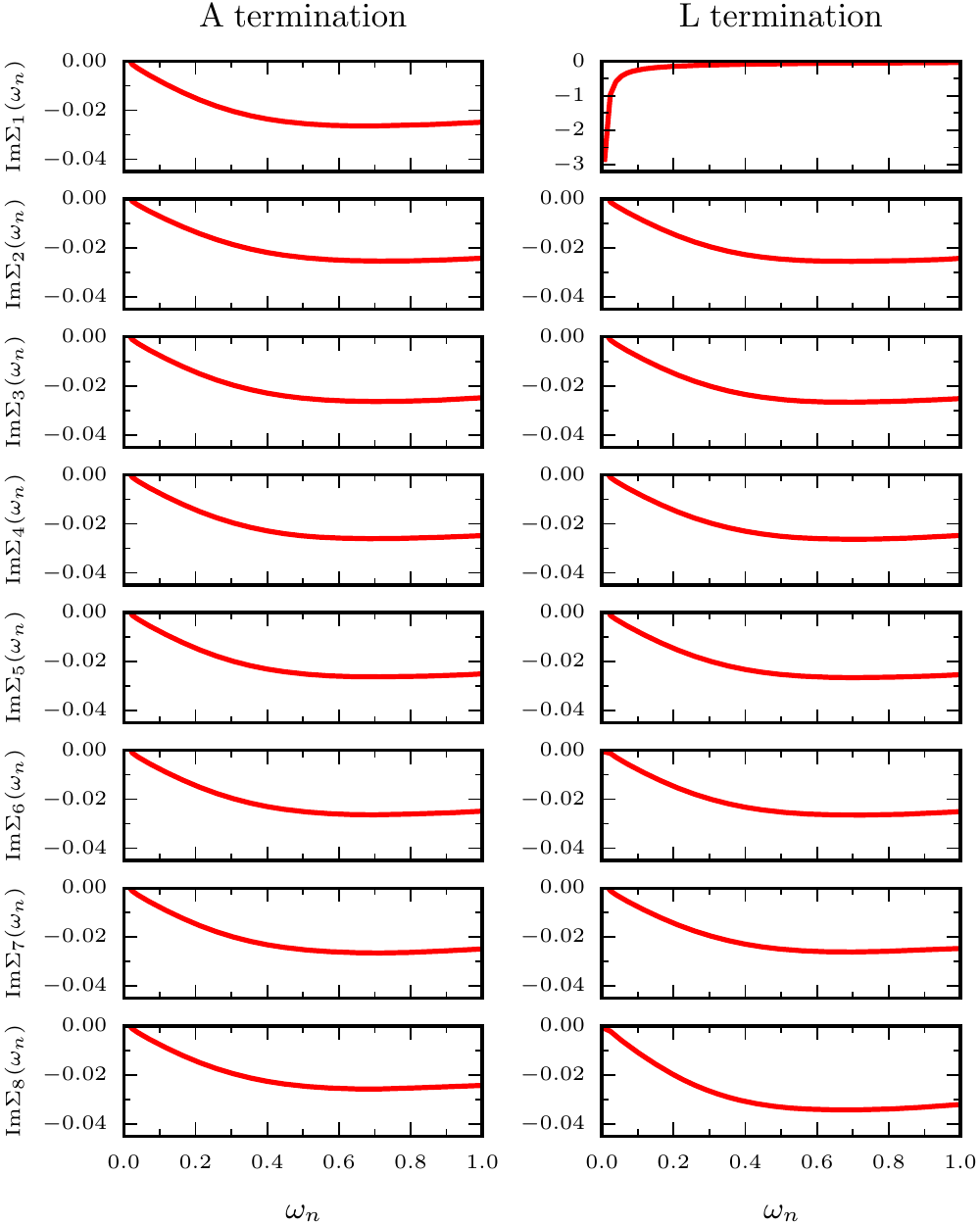}
    \caption{Imaginary parts of the impurity self-energies for the different layers and for the two terminations.}
    \label{Fig:Sigma}
\end{figure}

In Fig.~\ref{Fig:Sigma} we plot for all eight layers of the setup described in the main text the 
imaginary part of the impurity self-energy Im$\Sigma^{\mathrm{EDMFT}}_{\mathrm{aa}}\left(i\omega_n\right)$ 
on the Matsubara axis, as obtained from the solution of the coupled EDMFT impurity problems. The left column shows the results for the system with A termination and the right column those for the system with L termination. Only the surface layer of the
L-terminated system exhibits a Im$\Sigma$ which is diverging as $\omega_n\rightarrow0$, indicating
that the insulating state of the surface layer is due to correlations. In all the other layers  Im$\Sigma^{\mathrm{EDMFT}}_{\mathrm{aa}}\left(i\omega_n\right)$ is two orders of magnitude smaller and vanishes for $\omega_n\rightarrow 0$.

\newpage

\subsection*{Layer-resolved $\mathbf{k}_{\sslash}$-dependent spectral functions}
In Fig.~\ref{Fig:AkwLres} we show for the two terminations the layer-dependent spectral functions computed along 
a high-symmetry path within the Brillouin zone of the monolayer. The results were obtained with the maximum entropy method\cite{Jarrell1996} and
indicate that, when interactions are included, the distinctive feature of the surface layer
in the L-terminating system is the smaller gap. On the right side of each spectrum, we show the  $\mathbf{k}_{\sslash}$-integrated result. 
\begin{figure}[h]
    \centering
    \includegraphics[height=16cm]{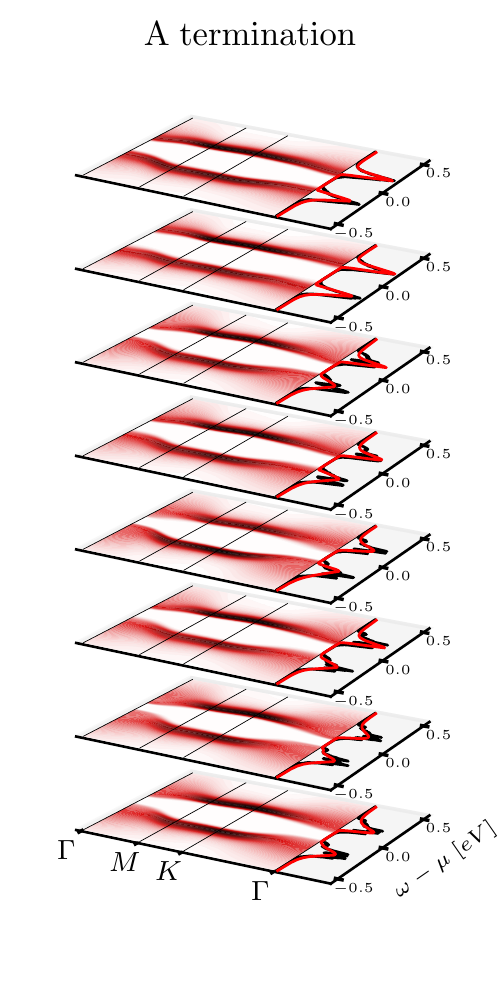}
    \includegraphics[height=16cm]{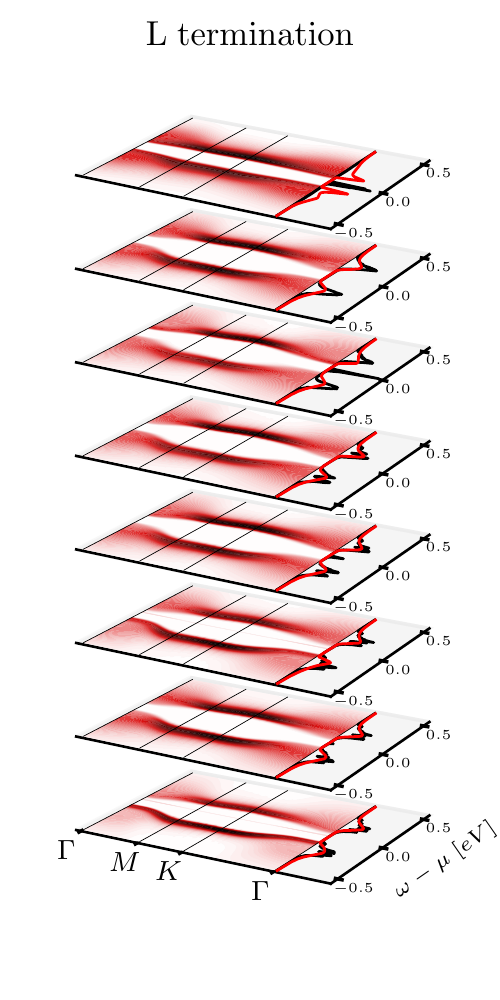}
    \caption{$\mathbf{k}_{\sslash}$-dependent spectral functions for each layer and for the two terminations. The non interacting surface spectral function for the L termination is rescaled by 0.2 as in the main text.}
    \label{Fig:AkwLres}
\end{figure}

\newpage

\subsection*{Experimental setup}
The line spectra in Fig. 3 of the main text were obtained using a commercial UV source (Harmonix, APE GmbH) generating tunable output in the range 5.7 – 6.3 eV. Harmonic generation in non-linear crystals was driven by the output of a tunable optical parametric oscillator pumped by a 532 nm Paladin laser (Coherent, inc.) at 80 MHz. The measurements presented here were obtained with 6.2 eV photons. The spectra were obtained by integrating the 2D ARPES data over a square area with linear extent $\pm 0.036~$\AA$^{-1}$ around the $\overline{\Gamma}$-point.  The sample surface was scanned by the encoded motion of a 6-axis cryogenic manipulator (SPECS GmbH). All spectra were acquired using a Scienta-Omicron DA30 analyzer.

\end{document}